\documentclass[prl,twocolumn,showpacs,preprintnumbers]{revtex4}
\usepackage{graphicx}
\newcommand{\Vec}[1]{\mbox{\boldmath$#1$}}
\begin{document}

\title{
Exact supersymmetry in the relativistic hydrogen atom in general dimensions\\
--- supercharge and the generalized Johnson-Lippmann operator
}
\author{
Hosho Katsura\cite{addr} and Hideo Aoki
}

\affiliation{Department of Physics, University of Tokyo, Hongo,
Tokyo 113-0033, Japan}

\date{\today}

\begin{abstract}
A Dirac particle in general dimensions moving in a $1/r$ potential is 
shown to have an exact ${\cal N} = 2$ supersymmetry, for which 
the two supercharge operators are obtained in terms of (a $D$-dimensional 
generalization of) the Johnson-Lippmann operator, an extension of the 
Runge-Lenz-Pauli vector that relativistically incorporates spin degrees of 
freedom. So the extra symmetry (S(2))
in the quantum Kepler problem, which determines the degeneracy of 
the levels, is so robust as to accommodate 
the relativistic case in arbitrary dimensions. 
\end{abstract}
\maketitle

{\it Introduction --- }
There is an increasing amount of fascination with supersymmetry (SUSY) 
in various fields of physics.\cite{SUSY,Physica}  
In a broad context SUSY is a kind of (graded) Lie albebra 
that closes under a combination of commutation and 
anticommutation.\cite{Cooper} 
One of the simplest realizations of such a symmetry may be found 
in one of the oldest problems in physics --- motion of an 
electron in a hydrogen atom, or, classically, Kepler's problem.  
It has long been known that there is a hidden symmetry (a dynamical symmetry) 
in the 
problem, which is related to an extra conservation (the Runge-Lenz vector).  
So the symmetry of the problem is higher (SO(4)) than the trivial SO(3), 
which is the cause of the ``accidental" degeneracy 
(i.e., the energy level independent of $l$) in the 
spectrum of the hydrogen in non-relativistic quantum mechanics. 
If we move on to the Dirac electron in hydrogen in the 
relativistic quantum mechanics, the accidental degeneracy 
is lifted (since the Runge-Lenz vector, which designates the 
direction of the perihelion, is no longer conserved).  
However, the degeneracy is lifted only incompletely, 
and two-fold degeneracies (for the Dirac-operator quantum number 
$\kappa \equiv -(2\Vec{S}\cdot \Vec{L} +1) = \pm(j+1/2$) 
with $j$ being the total angular momentum) remain, which was puzzling.  

In 1985 Sukumar made an interesting suggestion 
that the strange degeneracy may be 
explained as a supersymmetry in the problem\cite{Sukumar}.  
Subsequently however, Tangerman et al.,\cite{Tangerman} 
who have pointed out that there is indeed an exact ${\cal N} = 2$ 
supersymmetry for the {\it non}relativistic 
hydrogen atom, criticized that 
Sukumar's work has not actually constructed supercharges 
for the relativistic hydrogen atom.  
On the other hand, analytic studies for the relativistic 
hydrogen in general spatial dimensions have been developed, 
and analytic solutions are now obtained\cite{Gu,Dong}.  

Given this background, the purpose of the present Letter 
is to show that an exact ${\cal N} = 2$ 
supersymmetry for the relativistic Dirac particle 
in a $1/r$ potential in fact exists in general $D$ spatial dimensions.
In doing so we have actually constructed the supercharges $Q_{\pm}$ 
(i.e., mutually {\it anti}commuting operators that commute with the 
Hamiltonian) for the Dirac Hamiltonian in $(D+1)$ dimensions.  
The symmetry enables us to obtain the lowest eigenenergy 
and its wave function for each sector of (the $D$-dimensional 
generalization of) $|\kappa|$ in a simple and transparent manner, 
so the problem indeed turns out to be algebraically solvable.  

One interesting point is whether the supersymmetry in the hydrogen 
problem is related to the Runge-Lenz vector that 
describes the hidden conservation law on the nonrelativistic level.  
Dahl et al.\cite{Dahl} have shown, for the relativistic 
hydrogen in (3+1) dimensions, that 
the Runge-Lenz-Pauli vector (that necessarily involves 
spin degees of freedom for the Dirac particle, and is 
called Johnson-Lippmann operator\cite{Johnson}) may be indeed used to 
construct supercharges describing the supersymmetry.  
So the question we address here amounts to 
whether this extends to general dimensions.  
We first construct the Johnson-Lippmann operator generalized to 
the relativistic hydrogen in $(D+1)$ dimensions, 
and then show that the generalized operator 
may actually be used to construct the supercharges.  
The supercharges are in fact shown to reduce to 
(the $(D+1)$ dimensional generalization 
of) Runge-Lenz-Pauli vector in the nonrelativistic limit.  
While the Runge-Lenz-Pauli vector in general dimensions has been discussed 
in Refs.\cite{Sudarshan,Kirchberg} for the nonrelativistic case, 
the relativistic one is obtained for the first time.  
So we shall conclude that the supersymmetry is unexpectedly 
robust enough to be extended to the most general case, 
i.e., relativistic case in general dimensions. 
We can summarize the situation as\par

\begin{tabular}{l|l|l}
  & $D=3$ & general $D$ \\ \hline
nonrelativistic & SUSY\cite{Tangerman} & energy levels\cite{Dnonrel} \\ \hline
relativistic & SUSY\cite{Dahl} & SUSY (present work) \\ 
\end{tabular}

\noindent 
 
{\it Dirac's equation ---} 
Dirac's equation for the relativistic hydrogen atom in 
$(D+1)$ dimensions is written, in natural units (where 
$c=\hbar=1$), as
\begin{equation}
\{\gamma^{0}(p_0-eA_0)+\gamma^ip_i-m\} \Psi (x) = 0
\end{equation}
in standard notations, where $\gamma$'s satisfy 
$\{\gamma^{\mu},\gamma^{\nu}\}=g^{\mu \nu} 
= {\rm diag}(1,-1,-1,...)$, 
$p_{\mu}=i\partial_{\mu}$, $x=(x^0,x^1,\ldots,x^{D})$, 
and summations over repeated indices are implied.  
We assume a $1/r$ potential, so we have $eA_0=-Z\alpha/r$, 
where $\alpha=e^2/(\hbar c)$ is the fine-structure constant and 
$Z$ the atomic number.  When the lines of electric 
force are allowed to spread in the $D$ spatial dimensions 
the Coulomb potential becomes $1/r^{D-2}$, but we focus on 
the $1/r$ potential for general dimensions to retain the 
atomic stability\cite{Ehrenfest}.  We come back to this point 
later.  
If the time derivative is explicitly written, we have 
\begin{eqnarray}
i\frac{\partial}{\partial x^0}\Psi (x^0,x^1,\ldots,x^D) &=& H\Psi(x^0,x^1,\ldots,x^D),\\ \nonumber
H = \gamma^0m + \gamma^0\gamma^ip^i &-& \frac{Z\alpha}{r} \: (i=1,2,\ldots ,D)
\end{eqnarray}
where $H$ is the Hamiltonian and summations are implied for repeated spatial 
superscripts.  

The operators that commute with the Hamiltonian are the total 
angular momentum (in $D$ spatial dimensions), 
\begin{eqnarray}
J_{ab} &=& L_{ab}+\frac{i}{2}\gamma^{a}\gamma^{b}, \\ \nonumber
&& L_{ab}=ix_{a}\partial _{b}-ix_{b}\partial _{a}, 
\end{eqnarray}
and the Dirac operator (in $D$ dimensions), 
\begin{eqnarray}
K &\equiv& 
\gamma^{0}\left\{ \frac{i}{2}\sum_{a\neq b}^{} \gamma ^{a}\gamma ^{b}L_{ab}
+\frac{1}{2}(D-1) \right\} \\ \nonumber
&=& 
\gamma^{0}\left\{ \Vec{J}^2 -  \Vec{L}^2 - \Vec{S}^2 + \frac{1}{2}(D-1) 
\right\},
\end{eqnarray}
that is related to the spin-orbit interaction.

{\it Generalized Johnson-Lippmann operator ---} 
While $\Vec{J}$ and $K$ are (usually the only) 
constants of motion for arbitrary central 
fields, it is known, for the ordinary $D=3$, that 
there is an extra operator, a relativistic analogue of 
the Runge-Lenz-Pauli vector, that commutes with the Hamiltonian, as 
first constructed by Johnson and Lippmann\cite{Johnson}.  
We now generalize the Johnson-Lippmann operator 
to $(D+1)$ dimensions to 
define a Johnson-Lippmann-Katsura-Aoki operator,
\begin{equation}
A \equiv \gamma^{D+1}\gamma^{0}\gamma^{i}\frac{x^i}{r} 
- \frac{i}{Zm\alpha}K\gamma^{D+1}(H-\gamma ^{0}m).
\end{equation}
Here we have defined $\gamma^{D+1}$, a pseudo-scalar, which 
is a generalization of $\gamma^5$ in (3+1) dimensions, and which satisfies
$(\gamma^{D+1})^{\dagger} = \gamma^{D+1}, 
(\gamma^{D+1})^{2} = 1, 
\{  \gamma^{D+1}, \gamma^{\mu} \} = 0$.
$\gamma ^{D+1}$ is constructed from $\{ \gamma^{0},\gamma^{1},\ldots, 
\gamma^{D}\}$, but its actual form depends on whether the spatial 
dimension is even or odd.  

From the (anti)commutation relations above, we can show, after a 
rather tedious manipulation, that we have indeed 
$[H,A] = 0$ with a notable relation between $H$ and $A$,
\begin{equation}
A^{2} = 1 + \left(\frac{K}{Z\alpha}\right)^{2}
\left(\frac{H^2}{m^2}-1\right).
\label{A2}
\end{equation}
The expression reduces to the $D=3$ counterpart obtained in Refs.\cite{Dahl,Nikitin}.  

For $K$ and $A$, on the other hand, we can show that
\[
\{ A, K \} = 0,
\]
i.e., we end up with mutually anticommuting operators, 
$K, A$, that commute with $H$.  

{\it Construction of the ${\cal N}=2$ supercharge operators --- }
We are now in position to construct the generators of 
the supersymmetry, which was left undone in \cite{Dahl} and 
stated in Ref.\cite{Tangerman} as an operator interesting to find.  We can 
do so by going back to an ${\cal N}=2$ supersymmetric quantum 
mechanical model originally conceived by Witten\cite{Witten,Cooper}, to 
which the present Hamiltonian is shown to be formally equivalent.  
Due to the relation (\ref{A2}), we can take 
${\cal H} \equiv A^2$ as a Hamiltonian of the present problem.  
We can construct two operators $Q_{1},Q_{2}$ that commute with 
${\cal H}$, 
\begin{equation}
Q_1 = A, \; \; Q_2 = i\frac{AK}{|\kappa|},
\end{equation}
where $\kappa$ is the eigenvalue of $K$.  
We have then
\begin{eqnarray}
{\cal H} = Q_{1}^{2} &=& Q_{2}^{2}, \\ \nonumber
\{ Q_{1}, Q_{2} \} &=& 0.
\end{eqnarray}
If we further define 
\begin{equation}
Q_{\pm } \equiv \frac{1}{2}(Q_{1}\pm iQ_{2}) = 
\frac{1}{2}\left(1 \pm \frac{K}{|\kappa|}\right)A = 
\frac{1}{2}A(1 \mp {\cal P}_{\kappa})
\label{Qpm}
\end{equation}
with ${\cal P}_{\kappa} \equiv K/|\kappa|$, we have 
\begin{eqnarray}
Q_{+}^{2} = Q_{-}^{2} = 0, \\ \nonumber
{\cal H} = \{ Q_{+}, Q_{-} \}.
\end{eqnarray}
This establishes an equivalence to Witten's model.

So we can take the eigenvectors $|n,\pm \rangle$ that satisfy
\begin{eqnarray}
{\cal H} |n,\pm \rangle &=& E_{n}^{(\pm )}|n,\pm \rangle, \\ \nonumber
{\cal P}_{\kappa}|n,\pm \rangle &=& \pm |n,\pm \rangle ,
\end{eqnarray}
since ${\cal H}$$ = A^{2}$ commutes with $K$, and we can 
talk about the simultaneous eigenvectors.

${\cal P}_{\kappa}$ does not commute with $Q_{\pm }$, 
for which we have $Q_{\pm }|n,\pm\rangle=0 $. 
On the other hand, $[{\cal H}, Q_{\mp}]=0$ implies that 
$|n, \pm \rangle$ and $Q_{\mp}|n, \pm \rangle$ 
have a degenerate eigenvalue of ${\cal H}$ with 
different eigenvalues of ${\cal P}_{\kappa}$.  
From the relation (\ref{A2}) 
a zero-eigenstate of ${\cal H} = A^2$ 
is the ground state of the original $H$ 
(or, more precisely, the lowest-energy state for 
each value of $\kappa$).  
Since $\langle n, \sigma|{\cal H}|n,\sigma \rangle = 
\langle n, \sigma|\{ Q_{+},Q_{-}\}|n,\sigma \rangle 
= |Q_{+}|n,\sigma \rangle |^{2} + |Q_{-}|n,\sigma \rangle |^{2} \ge  0$ 
(for $\sigma = \pm$), a state is the ground state 
of $H$ if the equality holds in the above inequality.  The equality occurs when 
$Q_{+}|0,-\rangle = 0$ or $Q_{-}|0,+\rangle = 0$.
If we go back to the definition of $Q_{\pm}$ (eq.(\ref{Qpm})), 
the zero-eigenvalue state $|0 \rangle$ should satisfy
\[
A |0 \rangle = 0,
\]
since $(1{\mp}{\cal P}_{\kappa})|n,\mp\rangle \neq 0$.  

{\it Kernel of $A$ --- }
One way to establish the existence of such a 
state is an analytic method.  Following Ref.\cite{Gu}, 
we write the wave function for 
odd spatial dimensions $D = 2N + 1$ as
\begin{equation}
\psi_{\kappa }(x^{1},x^{2}, \ldots, x^{D}) = 
r^{-N}\left(
\begin{array}{@{\,}c@{\,}}
F(r)\phi_{\kappa}(\Omega) \\
iG(r)\phi_{-\kappa}(\Omega)
\end{array}
\right),
\end{equation}
where $F (G)$ is the ``large (small)" components, 
$\phi_{\kappa}$ is the angular part with angular coordinates 
$\Omega$, and 
$K\psi_{\kappa }=\kappa \psi_{\kappa }$ holds.  
If we plug this into $A\psi_{\kappa}=0$, we have, after some manipulation,
\begin{eqnarray}
\left[\left(x\frac{\,d}{\,dx}+{\rm sign}(\kappa)x\right)^{2}-s^{2}\right]
\left(\begin{array}{@{\,}c@{\,}}
f(x) \\
g(x)
\end{array}
\right) = 0 \\ \nonumber
s \equiv \sqrt{\kappa^{2}-(Z\alpha)^2},
\end{eqnarray}
where we have defined $F(r) \equiv f(x), G(r)\equiv g(x)$ 
in terms of a dimensionless $x \equiv (Z\alpha m/|\kappa|)r = (Z/|\kappa|a_0)r$ 
with $a_0$ being the Bohr radius.  

For this differential 
equation of second order, there are two independent solutions, 
$f(x) \propto 
x^{-s}e^{-{\rm sign}(\kappa)x}$ or 
$x^{s}e^{-{\rm sign}(\kappa)x}$ 
with 
\[
\kappa = \pm [l+\frac{1}{2}(D-1)], 
\]
where $l (=0,1,...)$ is the orbital angular momentum, 
but the only 
normalizable one is the latter with $\kappa>0$.  
With a similar argument for $g(x)$, we arrive at the kernel of $A$,
\begin{equation}
\psi_{\kappa} \propto x^{s-N}e^{-x}\left(
\begin{array}{@{\,}c@{\,}}
\phi _{\kappa }(\Omega) \\
i\frac{\kappa -s}{Z\alpha}\phi _{-\kappa}(\Omega)
\end{array}
\right)
\end{equation}
as the non-degenerate lowest state (the unpaired levels in Fig.1).  

\begin{figure}[t]
\begin{center}
\includegraphics[width=7.5cm,clip]{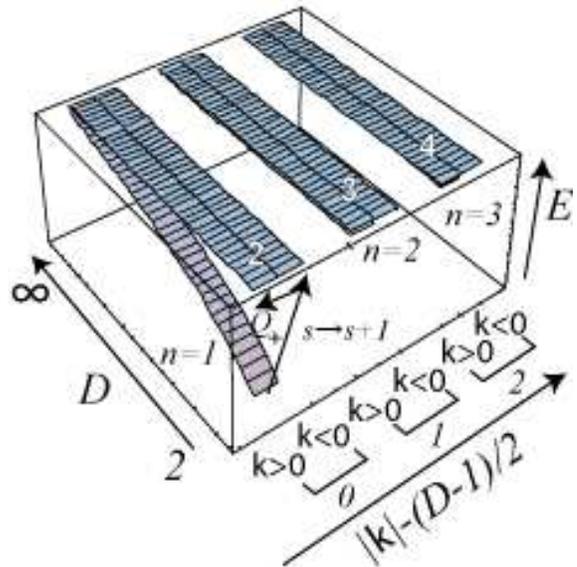}
\end{center}
\caption{Energy level scheme for the relativistic 
hydrogen in $D$ spatial dimensions.  The lowest few 
levels are plotted against the Dirac quantum number $\kappa$ 
and ${\rm tanh} D$, where the energy is plotted on a 
logarithmic scale to make the level splittings clearer.  
How the S(2) ladder is obtained with a shift in $s$ 
and the supercharge operation $Q_+$ is indicated by arrows with 
the sign of $\kappa$ and the principal quantum number $n$ indicated.
}
\label{fig1}
\end{figure}

For even $D = 2N$ we can perform a similar procedure, 
again following Ref.\cite{Gu}.  This time we can put 
$\psi_{\kappa} = r^{-N+1/2}[F(r)\phi_{\kappa}(\Omega)+
iG(r)\phi_{-\kappa}(\Omega)]$, 
for which the same differential equation results 
for $F(r), G(r)$, so we have a non-degenerate ground state, 
$\psi_{\kappa}
\propto 
x^{s-N+1/2}e^{-x}\{\phi_{\kappa}(\Omega)+
i[(\kappa-s)/Z\alpha]\phi_{-\kappa}(\Omega)\}$.  

{\it Eigenenergies and the group theory --- } 
Eigenenergies may readily be obtained algebraically: 
As used above, eq.(\ref{A2}) 
dictates that each of the 
zero-eigenstates of the supersymmetric Hamiltonian ${\cal H} = A^2$ 
is the lowest-energy state (for each sector of $K$) 
of the original $H$.  This immediately implies that 
$E$, the lowest-lying (within each sector of $K$) eigenvalue of $H$, is
\begin{equation}
E/m = \sqrt{1 -\left(\frac{Z\alpha}{\kappa}\right)^2} 
= \left[1 + \left(\frac{Z\alpha}{s}\right)^2\right]^{-1/2}, 
\end{equation}
in agreement with the analytical result in Ref.\cite{Gu}.  
We can go up the ladder (where the leg corresponds to 
$\kappa = \pm |\kappa|$ while the rung 
spanned by a quantum number we call $n'$) 
by making $s \equiv \sqrt{\kappa^{2}-(Z\alpha)^2} \rightarrow s+1$.  

So we end up with
\[
E/m = 
\left[1 + \frac{{(Z\alpha)^2}}
{\left(n-|\kappa|+\frac{D-3}{2}+\sqrt{\kappa^2-(Z\alpha)^2}\right)^2}\right]^{-1/2}, 
\]
where we assume $Z\alpha < (D-1)/2$ for the atomic stability.  
Here the  principal quantum number is given as
\[
n = l+1+n'
\]
with $n'=0,1,..,$ where each level is doubly degenerated 
[corresponding to $\kappa = \pm (l+(D-1)/2)$ and related by $Q_+$], except 
at the bottom of each ladder 
at $n'=0$ (i.e., $n=l+1$) for which only 
$\kappa >0$ should be taken (Fig.1).  
We can also see that the inter-dimensional degeneracy, 
which exists between the levels $(l,D) \rightarrow (l\pm1, D\mp2)$ 
noted for the nonrelativistic case\cite{Dnonrel}, 
persists for the relativistic case, and 
the supersymmetric ladders live on such a spectrum. 

Group-theoretically the present result implies the following.  
The nonrelativistic hydrogen atom in $D$ spatial dimensions 
has a hidden symmetry (with 
the Runge-Lenz vector conserved), and the symmetry of the 
problem is higher (SO($D+1$)) than the symmetry (SO($D$)) 
of the space.  If we go to the relativistic case the symmetry 
is degraded, but only partially degraded into SO($D$)$\otimes$S(2) 
due to the ${\cal N}=2$ supersymmetry.  
So the conjecture, stated in Ref.\cite{Tangerman}, is 
established here for general dimensions.  

{\it Relation with the Runge-Lenz-Pauli vector --- }
Related to the above, we can note a certain relation 
between the Johnson-Lippmann-Katsura-Aoki operator ($A$) and 
the Runge-Lenz-Pauli vector as follows.  
Since $A$ is Hermitian, we have 
\begin{eqnarray}
&A& = \frac{1}{2}(A+A^{\dagger}) 
= - \gamma ^{D+1}\gamma^{0}\gamma^{i} \\ \nonumber
&\times& \left[\frac{1}{2Zm\alpha}\gamma^{0}
(p^{j}L_{ij}-L_{ij}p^{j})-\frac{x^{i}}{r}\right] + 
\frac{i}{m}K\gamma^{D+1}\frac{1}{r}.
\end{eqnarray}
This expression reduces, in the nonrelativistic limit, to
\[
A \rightarrow -\sigma^{i}M^{i},
\]
where 
\[
M^{i}=\frac{1}{2mZ\alpha}(p^{j}L_{ij}-L_{ji}p^{j})-\frac{x^{i}}{r}
\]
is the nonrelativistic Runge-Lenz-Pauli vector 
in $D$ spatial dimensions\cite{Sudarshan,Kirchberg} 
and $\sigma^{i} = \gamma^{D+1}\gamma^{0}\gamma^{i}$ 
the spin operator in $D$ dimensions.   
Namely, the operator in this limit is the inner 
product of the Runge-Lenz-Pauli vector ($\Vec{M}$) 
and the spin ($\Vec{\sigma}$) in $D$ dimensions\cite{JL}.  

{\it Summary and discussions --- } 
So the higher symmetry for the $1/r$ potential in general dimensions 
is revealed to be surprisingly robust against 
the relativistic generalization and is retained as a supersymmetry.  
The supercharge is indeed related to (a $D$-dimensional 
generalization of) the Johnson-Lippmann operator.  

Given the formula for the general $D$, we are tempted to ask what 
happens in the 
$D \rightarrow \infty$ limit.  In the nonrelativistic case the kinetic 
energy ($\sim (1/D)\times$ potential energy) is dominated by the 
potential energy, so the system reduces to a set of harmonic 
oscillations around the classical potential minima as stressed 
in Ref.\cite{Dnonrel}.  
In the relativistic case, however, we see that we cannot eliminate 
the small component ($G$) in the 
$D \rightarrow \infty$ limit, since the coupling between $F$ and $G$ 
does not vanish in this limit.  This implies that we cannot trivially 
relate the supersymmetry with a set of oscillators even asymptotically.

Another basic question is whether the supersymmetry is 
just accidental to the $1/r$ potential.  
As mentioned above, electromagnetically there is a problem of 
how we can conceive the $1/r$ potential as the Coulomb 
potential in general dimensions.  We should have 
$1/r^{D-2}$ potential from Gauss's law if the lines of electromagnetic 
force extend over the $D$ spatial dimensions, but there is a 
well-known Ehrenfest's 1920 result that the atom 
becomes unstable for this potential.  Some authors 
argue that we should in fact stick to $1/r$ in general 
dimensions.\cite{Burgbacher}  
Experimentally, a low-dimensional ($D=2$) case may be 
interesting as an accessible one.  

In a broader context, a supercharge is a kind of generalization of 
the Dirac 
operator, so it could be that 
supersymmetry is shared by a wide class 
of Dirac-type equations.  Some authors\cite{Leach} have indeed discussed 
generalization of the 
Runge-Lenz-Pauli vector, such as the 
case of a particle around a magnetic monopole with a vector potential 
around it.  
So the exploration of supersymmetry in wider gauge-field models 
will be an interesting future problem.  
We wish to thank Dr R. Arita for discussions.


\begin{references}
\bibitem[*]{addr} Present address: Department of Superconductivity, 
University of Tokyo, Hongo, Tokyo 113-8656, Japan

\bibitem{SUSY} Steven Weinberg: {\it Supersymmetry} 
(Cambridge Univ. Press, New York, 2000).  

\bibitem{Physica} V.A. Kostelecky and D.K. Campbell (ed.): 
{\it Supersymmetry in Physics} (North-Holland, 1985).

\bibitem{Cooper} 
For a review on the supersymmetric quantum mechanics, see 
F. Cooper, A. Khare, and U. Sukhatme, 
Phys. Rep. {\bf 251}, 267 (1995). 

\bibitem{Sukumar} C.V. Sukumar, J. Phys: Math. Gen. {\bf 18}, L697 (1985). 

\bibitem{Tangerman} R.D. Tangerman and J.A. Tjon, Phys. Rev. A {\bf 48}, 
1089 (1993). 

\bibitem{Gu} X.Y. Gu, Z.Q. Ma, and S.H. Dong, Int. J. Mod. Phys. E {\bf 11}, 
335 (2002). 

\bibitem{Dong} S.H. Dong, G.H. Sun, and D. Popov, J. Math. Phys. {\bf 44}, 
4467 (2003).

\bibitem{Dahl} J.P. Dahl and T. Jorgensen, Int. J. Quantum Chem. {\bf 53}, 
161 (1995). 

\bibitem{Johnson} M.H. Johnson and B.A. Lippmann, Phys. Rev. {\bf 78}, 
329(A) (1950); see also V.B. Berestetskii et al., {\it Quantum Electrodynamics} 
(Pergamon, 1982). 

\bibitem{Sudarshan} E.C.G. Sudarshan, N. Mukunda, and L. O'Raifeartaigh, 
Phys. Lett. {\bf 19}, 322 (1965). 

\bibitem{Kirchberg} A. Kirchberg, J.D. Laende, P.A.G. Pisani, and 
A. Wipf, Annals Phys. {\bf 303}, 359 (2003).


\bibitem{Witten} E. Witten, Nucl. Phys. B {\bf 188}, 513 (1981). 

\bibitem{Dnonrel} For the nonrelativistic hydrogen atom in 
$(D+1)$ dimensions, there are analytic works 
[D.R. Herrick, J. Math. Phys, {\bf 16}, 281 (1975); 
D.R. Herschbach in {\it Atomic Physics} {\bf 11}, ed. by S. Haroche et al. 
(World Scientific, 1989), p. 63], 
along with algebraic approaches exploiting the 
Runge-Lenz-Pauli vector\protect\cite{Kirchberg}. 

\bibitem{Ehrenfest} P. Ehrenfest, Ann. Phys. {\bf 61}, 440 (1920).

\bibitem{Nikitin} A.G. Nikitin 
in {\it Photon and Poincar\'{e} Group} ed. by 
V.V. Dvoyeglazov (Nova Science, New York, 1999) p. 74.

\bibitem{JL} The result agrees 
with the nonrelativistic limit for $D=3$ discussed in Ref.\protect\cite{Dahl}.  
In this limit, the supercharge $Q_{1}, Q_{2}$ given in terms of $A$ 
here reduces to those defined in Ref.\protect\cite{Tangerman}.  

\bibitem{Burgbacher} 
F. Burgbacher, C. L\"{a}mmerzahl, and A. Macias, 
J. Math. Phys. {\bf 40}, 625 (1999). 

\bibitem{Leach} P.G.L. Leach and G.P. Flessas, J. Nonlinear Math. Phys. 
{\bf 10}, 340 (2003).

\end{references}
\end{document}